\documentclass[%
preprint,
 amsmath,amssymb,
 aps,
]{revtex4-2}

\usepackage{graphicx}
\usepackage{dcolumn}
\usepackage{bm}
\usepackage{xcolor}

\begin{document}

\preprint{APS/123-QED}

\title{A hidden mechanism of dynamic LES models}


\author{Xiaohan Hu}
\affiliation{  Department of Mechanical Engineering and Applied Mechanics, University of Pennsylvania, Philadelphia, PA 19104, USA }

\author{Keshav Vedula}%
\affiliation{Aerothermal Engineering Group, SpaceX, Hawthorne, CA 90250, USA}

\author{George Ilhwan Park}%
\email{gipark@seas.upenn.edu}
\affiliation{%
 Department of Mechanical Engineering and Applied Mechanics, University of Pennsylvania, Philadelphia, PA 19104, USA  }

%




\date{\today}

\begin{abstract}
The dynamic model is one of the most successful inventions in subgrid-scale (SGS) modeling as it alleviates many drawbacks of the static coefficient SGS stress models. The model coefficient is often  calculated dynamically through the minimization of the Germano-identity error (GIE). However, the driving mechanism behind the dynamic model's success is still not well understood. In wall-bounded flows, we postulate that the principal directions of the resolved rate-of-strain tensor play an important role in the dynamic models. Specifically, we find that minimization of the GIE along only the three principal directions (or less), in lieu of its nine components in its original formulation, produces equally comparable results as the original model when examined in canonical turbulent channel flows, a three-dimensional turbulent boundary layer,  and a separating flow over periodic hills. This suggests that not all components of the Germano identity are equally important for the success of the dynamic model, and that there might be dynamically more important directions for modeling the subgrid dynamics. 
\end{abstract}

\maketitle


\section{\label{sec:intro}Introduction}
Dynamic closure of the subgrid-scale (SGS) stress is perhaps the most celebrated feature of large-eddy simulations (LES), which is absent in other lower-fidelity approaches for modeling turbulence. The class of dynamic SGS models allows for the determination of model coefficients purely from the resolved-scale information available in live LES calculations,  eliminating the need for the ad-hoc practice of parameter calibration and therefore greatly promoting the predictive nature of the method. It was \citet{Germano1991} who first introduced the idea of the dynamic procedure,  proposing the dynamic Smagorinsky model (DSM). This formulation was based on the Germano identity which is an algebraic relation between the SGS stresses at two different filter levels and the resolved turbulent stresses. \citet{Lily1992} proposed a modification by minimizing the Germano-identity error (GIE) which has become the most widely adopted practice. 

Compared to the massive works that apply the DSM to the study of turbulent flows, the mechanism behind why the DSM is successful is much less understood.
The explanation based on scale invariance in the inertial subrange was initially adopted, but it was later challenged by \citet{Jimenez2000} who claimed that the DSM's success is thanks to the model's robustness to errors in the physics. 
\citet{Pope2004} brought another perspective that the dynamic procedure minimizes the dependence of relevant turbulence statistics (total Reynolds stresses) on filter levels. \citet{Toosi2021} complemented the understanding by showing the connection between the GIE and the residual of the LES governing equations.

Closely related to the interpretation of the GIE tensor, how one minimizes the GIE can make a difference in the SGS stress modeling. \citet{Ghosal1995} recast the solution procedure of the model coefficient in the context of a variational problem, generalizing the dynamic procedure to flows without homogeneous directions. 
\citet{Meneveau1996} introduced a Lagrangian dynamic procedure where the GIE is minimized along the flow pathlines, allowing for the application of the model to inhomogeneous flows in complex geometries.  
\citet{Morinishi2001} proposed a modification for the dynamic two-parameter mixed model to improve the model performance in wall-bounded turbulent flows. \citet{NomaPark2009} explored reduction in an ensemble-averaged GIE and proposed an efficient predictor-corrector-type method to find the optimal parameter. \citet{Denaro2013} derived the integral-based Germano identity which showed much less sensitivity to the type of contraction than expected in the differential-based formulation. \citet{Agrawal2022} proposed a tensorial Smargorinsky coefficient in the DSM to overcome the invalid assumption of alignment between the filtered strain-rate tensor and the SGS stress.

The present study aims to provide an alternative explanation of the mechanism 
behind some dynamic SGS models rooted in the Germano identity, focusing on wall-bounded flows. Specifically, we show 
that only a few directions matter for these models, namely, the principal directions of the filtered strain-rate tensor. The idea 
is demonstrated in turbulent channel flow, a three-dimensional turbulent boundary layer (3DTBL),  and a separating flow over periodic hills. The manuscript is organized as follows. In section \ref{sec:mod},  some reduced dynamic procedures 
designed to highlight characteristic behaviors of the dynamic model along these directions 
are presented. Three flow configurations used in the present work are explained in section \ref{sec:flow}. Section \ref{sec:result} presents the main results and analyses, which is followed by the conclusion in Section \ref{sec:conclusion}. 

\section{\label{sec:mod}
Reduction of dynamic procedures along the principal directions of $\bar S_{ij}$}
We first summarize the standard dynamic procedure deployed widely in many  dynamic models. 
In Smagorinsky-type models, the deviatoric part of the SGS stress 
tensor $\tau_{ij}$ is modeled as 
\begin{equation} \label{eq:Smagorinsky}
    \tau_{ij} - \frac{1}{3}\delta_{ij}\tau_{kk} = 2C\Delta^2|\overline{S}|\overline{S}_{ij}, 
\end{equation}
where $\delta_{ij}$ is the Kronecker delta, 
$\Delta$ is the grid filter size, 
$\overline{S}_{ij} = \frac{1}{2}\left ( \frac{\partial \overline{u}_i}{\partial x_j}+\frac{\partial \overline{u}_j}{\partial x_i}\right )$ 
is the resolved strain-rate tensor at the grid filter level, 
and $|\overline{S}| = \left (2\overline{S}_{kl}\overline{S}_{kl}\right )^{1/2}$.  
The overbar $\ \bar{\cdot}\ $ denotes the grid-filtered quantities. 
The Smagorinsky coefficient $C$ is determined by a dynamic procedure \citep{Lily1992} based on the Germano identity (GI), 
\begin{equation} \label{eq:Germano_identity}
    L_{ij}  = T_{ij} - \widehat{\tau_{ij}}, 
\end{equation}
where $T_{ij}$ is the SGS stress at the test-filter level defined as 
\begin{equation} \label{eq:test stress}
    T_{ij}  = \widehat{\overline{u_i}}\ \widehat{\overline{u_j}} - \widehat{\overline{u_iu_j}}, 
\end{equation}
and $\tau_{ij}$ is the SGS stress at the grid-filter level
\begin{equation} \label{eq:grid stress}
    \tau_{ij}  = \overline{u_i}\ \overline{u_j} - \overline{u_iu_j}.    
\end{equation}
The overhat $\ \widehat{\cdot}\ $ denotes the test-filtered quantities. 
$L_{ij}$ contains the resolved components of the stress tensor associated with scales between the test and grid filter scales, and it can be computed directly from 
the information available in the LES calculations, 
\begin{equation} \label{eq:lij}
    L_{ij} = -\widehat{\overline{u_i}\ \overline{u_j}}+\widehat{\overline{u_i}}\ \widehat{\overline{u_j}}, 
\end{equation}
using Eq.~(\ref{eq:Germano_identity})$-$(\ref{eq:grid stress}).
$T_{ij}$ is modeled similarly as in Eq.~(\ref{eq:Smagorinsky}), 
using the same model coefficient $C$ under the scale-invariance ansatz. 
Substitution of the modeled stresses into the
deviatoric part of the 
Germano identity 
produces an over-determined system for the unknown coefficient $C$, 
\begin{equation} \label{eq:Germano}
    L_{ij} - \frac{1}{3}\delta_{ij}L_{kk} = CM_{ij}        
\end{equation}
where
\begin{equation} \label{eq:mij}
    M_{ij} = 2\widehat{\Delta}^2 |\widehat{\overline{S}}|\widehat{\overline{S}}_{ij} - 2\Delta^2\widehat{|\overline{S}|\overline{S}_{ij}}
\end{equation}
is again computable with the LES solution. 
Here, $\hat \Delta$ is the test-filter size typically taken as $\hat \Delta = 2 \Delta$.  
The commonly used procedure is the least squares approach, which minimizes the $L_2$ norm of the GIE tensor, $Q = Q_{ij}Q_{ij}$ \citep{Lily1992}.
Here, $Q_{ij}$ is the GIE tensor defined as the residual of the Germano identy 
\begin{equation} \label{eq:Germano error}
    Q_{ij} = L_{ij} - \frac{1}{3}\delta_{ij}L_{kk} - CM_{ij},   
\end{equation}
 and the coefficient $C$ is then determined through a least-square procedure  as 
\begin{equation} \label{eq:dsml2}
    C = \left < L_{ij}M_{ij}\right > 
    \Big / 
    \left < M_{ij}M_{ij}\right >.      
\end{equation}
Here,  $\left<\cdot\right>$ denotes the averaging in homogeneous directions (if any) or local filtering operation used to stabilize the model. 
This original 
dynamic procedure  accounts for all components (and therefore directions) of the GIE tensor collectively with equal weights. 

Vorticity dynamics in the inviscid limit implies that vortices are frozen to fluid elements and therefore they deform in the same way fluid elements do. 
As the strain-rate tensor characterizes the local deformation state of fluid elements, vortices are more likely aligned with the principal directions of the strain-rate tensor \citep{davidson2015turbulence, misra1997vortex}. 
If one adopts the scale-similarity ansatz \citep{bardina1980improved},  it can be further assumed that the most energetic SGS eddies are oriented primarily by the smallest resolved scale. Motivated from this line of argument, 
we postulate 
that there are a few dynamically more important directions
which embody the essence of the dynamic procedure, namely, the principal directions of the resolved strain rate field. 
Numerical experiments with the dynamic procedures further reduced along these directions can 
be used to test this idea. 
To this end, 
we focus  on satisfying the GI along the principal directions of $\bar S_{ij}$ only, and examine effectiveness of this hypothesis. 
Three closely related formulations for this purpose are introduced below. 

Dynamic procedures which account for the GI along the principal directions of $\overline{S}_{ij}$ only can be expressed in a general form as 
\begin{equation} \label{eq:pd_general}
    C = \sum\limits_{j=1}^n  \left < 
    \alpha_j L'_{jj}M'_{jj}\right >
    \Big / 
    \sum\limits_{j=1}^n \left < 
    \alpha_j M'_{jj}M'_{jj}\right >, 
\end{equation}
where $\alpha_j$ are proper weights for the $j^{th}$ principal direction, 
and the prime symbol ($'$) is used to denote tensors represented in the eigen coordinate of $\overline{S}_{ij}$. For instance, 
$Q'_{kl} = V^{-1}_{ki}Q_{ij}V_{jl}$, where $V_{ij}$ contains the orthonormal eigenvectors  of $\overline{S}_{ij}$. 
The first formulation denoted as PDL2 ($L_2$ norm minimization along principal directions) is defined as $n=3$  and $\alpha_j = 1$. This approach minimizes the modified cost function $Q = \sum\limits_{j=1}^3 (Q'_{jj})^2$, i.e., the squared sum of the GIE along the principal directions of $\overline{S}_{ij}$. 
The second formulation denoted as PDWL2 is defined as $n=3$ 
and $\alpha_j = \lambda_j^2$, where $\lambda_j$ are the eigenvalues of $\overline{S}_{ij}$. 
This approach minimizes $Q = \sum\limits_{j=1}^3 (\lambda_j Q'_{jj})^2$, i.e., the squared sum of the GIE weighted according to the level of stretching/compression along the principal directions of $\overline{S}_{ij}$. 
A maximally reduced version is where $n=1$ and $\alpha_j = 1$, which cares only about the direction with the maximum stretch: $C$ is determined from the GI applied along the direction with the maximum positive eigenvalue of $\overline{S}_{ij}$. This approach (denoted as PDMAX) assumes that the SGS eddies align along 
 the maximal vortex stretching direction of the resolved-scale eddies, and only that direction matters to the SGS dynamics/energetics. 
 It should be noted that the coordinate invariance of Eq.~(\ref{eq:pd_general})
 is guaranteed from  the fact 
 the eigenvalues and eigen directions of a tensor are invariant in any coordinate system. 
 The eigen coordinate system of $\overline{S}_{ij}$ is unique at the moment 
Eq.~(\ref{eq:pd_general}) is to be evaluated, and 
 any tensor's representation in this coordinate system is also unique.

 \section{\label{sec:flow}Flow configuration}
The first case considered in the present work is the plane turbulent channel flow with periodic boundary conditions in the streamwise and spanwise directions. 
DNS results from \citet{Moser1999} and the Johns Hopkins Turbulence Database (JHTDB) \citep{JHTDB1,JHTDB2} are used as reference.
The computational domain is set to be $(L_{x},L_{y},L_{z})=(2\pi\delta,2\delta,2\pi\delta/3)$ for $Re_{\tau} = 395$ and $(L_{x},L_{y},L_{z})=(2\pi\delta,2\delta,\pi\delta)$ for $Re_{\tau} = 1000$, where $x$ is the streamwise direction, $y$ is the wall-normal direction and $z$ is the spanwise direction. $\delta$ is half channel height. The flow is driven by the constant pressure gradient in the streamwise direction. 

The second case is the three-dimensional boundary layer created on a flat plate by a time-dependent freestream velocity vector, whose magnitude is independent of time but whose direction changes at a constant angular velocity \citep{Spalart1989}. The Reynolds number ($Re_l = U_0\left(\frac{2}{f\nu}\right)^{1/2}$) is 767. Here, $U_0$ is freestream velocity magnitude, $f$ is the angular rate of rotation of the freestream velocity vector and $\nu$ is the kinematic viscosity.
    In our numerical simulation, the computational domain is set to be $(L_{x},L_{y},L_{z})=(2\delta,\delta,2\delta)$, where $y$ is the wall normal direction. $\delta=\frac{u^*}{f}$ is the outer length scale where $u^*$ is the velocity scale as defined in \citet{Spalart1989}. The top boundary condition is set to be the rotating velocity vector,
    \begin{equation}
        U_{\infty} = U_{0} \cos(ft),\quad W_{\infty} = U_{0} \sin(ft). 
    \end{equation}
    Periodic boundary conditions are applied to the two horizontal directions, $x$ and $z$.
    Despite its simple configuration, the flow is characterized with a skewed mean velocity profile (i.e., the flow direction varying with the wall distance) and a full Reynolds-stress tensor, similar to the Ekman layer.  
    The flow statistics are computed in the coordinate system that is rotating with the freestream velocity vector. In this coordinate system, the flow is statistically steady. 

The third case is the separating flow over periodic hills \citep{Rapp2011}. The computational domain is $(L_{x},L_{y},L_{z})=(9h,3.035h,4.5h)$ where $h$ is the height of the hill. $x$, $y$ and $z$ denote the streamwise, wall-normal and spanwise directions respectively. The Reynolds number based on the hill height $h$ and bulk velocity above the hill crest $U_S$ is $Re_S = U_Sh/\nu$. It is related to the domain-averaged bulk Reynolds number ($Re_B = U_Bh/\nu$) by a factor of 0.72, $Re_B = 0.72Re_S$. The flow is driven by a constant mass flow rate. Periodicity is applied to the streamwise and spanwise directions.

\section{\label{sec:result}Results and discussions}
The simulations are performed with CharLES, an unstructured cell-centered finite-volume compressible LES solver developed at Cascade Technologies, Inc. The solver employs an explicit third-order Runge-Kutta (RK3) scheme for time advancement and a second-order central scheme for spatial discretization. More details regarding the flow solver can be found in \citet{Khalighi2011} and \citet{Park2016}.

\subsection{Turbulent channel flow at $Re_{\tau} =395$ and $Re_{\tau} =1000$}
\begin{figure}
\centerline{\includegraphics[width=1\linewidth]%
{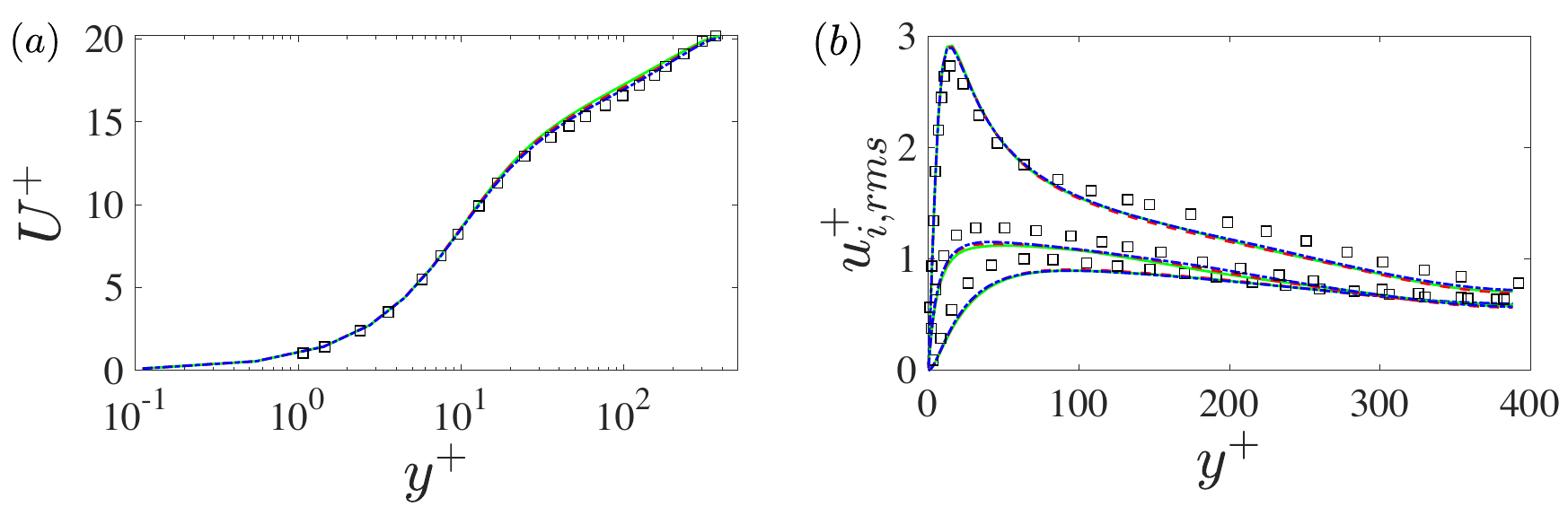}}
\caption{Profiles of flow statistics in wall units in turbulent channel flow at $Re_{\tau}=395$. ($a$) Mean streamwise velocity; ($b$) Turbulence intensities ($u_{rms}$, $v_{rms}$ and $w_{rms}$). Black squares, DNS \citep{Moser1999}; green solid line, LES with DSM; red dashed line, LES with PDL2; blue dash-dotted line, LES with PDWL2.}
\label{fig:mean_channel395}
\end{figure}

\begin{figure}
\centerline{\includegraphics[width=0.7\linewidth]{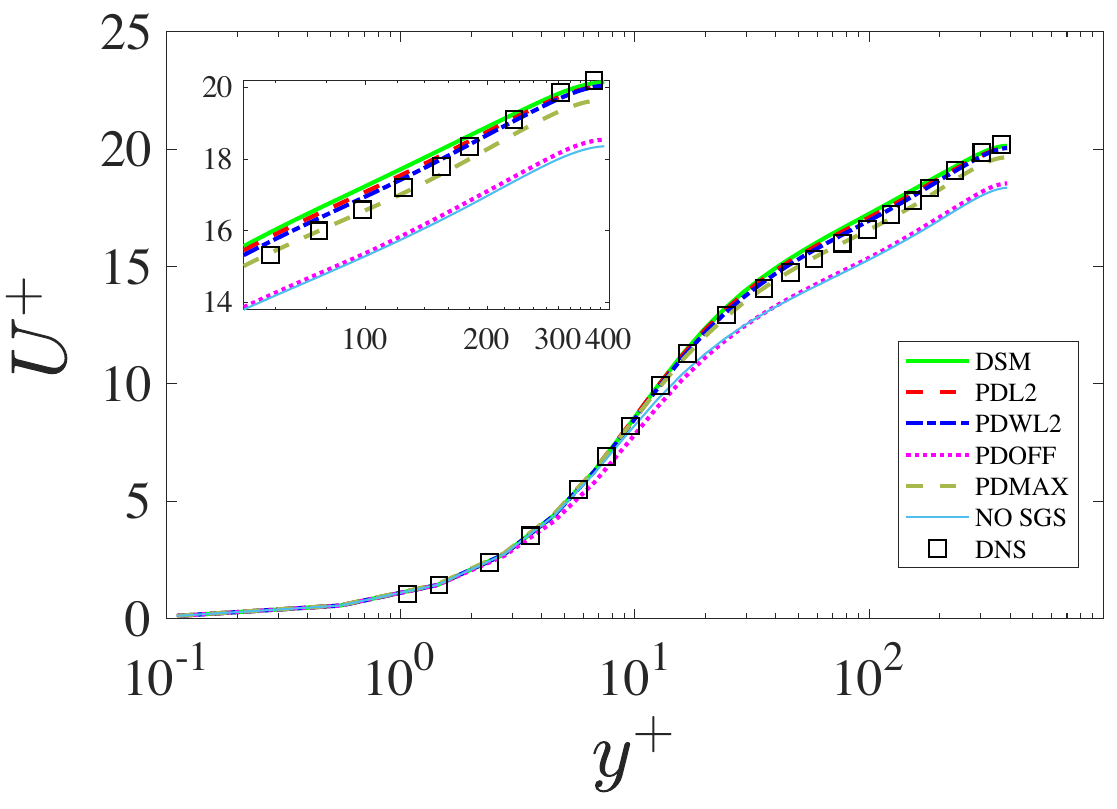}}
\caption{Mean velocity profiles of  channel flow at $Re_{\tau}=395$. Squares, DNS \citep{Moser1999}; green solid line, LES with DSM; red dashed line, LES with PDL2; blue dash-dotted line, LES with PDWL2; magenta dotted line, LES with non-principal components model, PDOFF; olive green dashed line, LES with PDMAX; cyan solid line, no SGS model.}
\label{fig:mean_channel_pdoff}
\end{figure}

Figure~\ref{fig:mean_channel395} shows the profiles of flow statistics for turbulent channel flow at $Re_{\tau}=395$. The grid spacing in wall units 
is $(\Delta_x^+,\Delta_y^+,\Delta_z^+)=(50,0.22 \sim 13,16.5)$. 
The LES results agree well with DNS \citep{Moser1999} in terms of mean velocity. As commonly reported in underresolved LES \citep{bae2018turbulence},  a slight overprediction of the streamwise intensity ($u_{rms}$) 
and underprediction of the other intensities are observed. 
All dynamic procedures (the original and PD versions from Sec.~\ref{sec:mod}) produce nearly identical result, 
but the PD versions are seen slightly more accurate when zoomed in (see  Fig.~\ref{fig:mean_channel_pdoff}). 
PDL2 and PDWL2 which use only three diagonal components of the GIE tensor in the principal coordinate system of the grid-filtered strain-rate tensor perform equally well compared to the original DSM, which includes all components of the GIE tensor. It should be noted that all components of the GIE tensor in the eigen coordinate of $\overline{S}_{ij}$ were found to be nonzero and comparable in their magnitude. 
The results here imply that not all the components of the GI are equally important. By working on only partial information of the GI, the dynamic model can produce almost identical results to the original DSM results.
Although not shown here for brevity, an identical behavior was observed in a channel flow calculation with $Re_{\tau}=1000$ using 
a relatively coarser grid with $(\Delta_x^+,\Delta_y^+,\Delta_z^+)=(100,0.5 \sim 32,50)$. 

To further highlight how different components of the GIE tensor contribute to the performance of the DSM, another two reduced dynamic procedures are tested in the same turbulent channel flow at $Re_{\tau}=395$ as a comparison. The first one includes only the non-principal components (off-diagonal components of the GIE tensor represented in  the principal coordinates of $\overline{S_{ij}}$) in the dynamic procedure. The model is referred to as PDOFF. Another model is
PDMAX introduced earlier in Sec.~\ref{sec:mod}, which operates only on the principal direction of $\overline{S_{ij}}$ with 
the maximum stretching.  
The mean velocity profiles are shown in Fig.~\ref{fig:mean_channel_pdoff}. It can be observed that PDOFF underpredicts the mean velocity, and interestingly, it performs as bad as the no SGS model result. This indicates that the non-principal components have no contribution in the determination of the eddy viscosity. On the other hand, the PDMAX model performs similarly as the original DSM. It has a better agreement with DNS between $50<y^+<110$ but slightly underpredicts the mean velocity for $y^+>110$. The reasonably good performance of the PDMAX model is especially surprising given it only considers one component of the GI tensor. This may imply that the SGS model can be further reduced, and the eddy-stretching directions are potentially more important than the eddy-compressing directions. However, it should be noted that clipping to avoid negative eddy viscosity
was necessary for the PDMAX version above  $y^+=100$ (which can explain its underprediction for $y^+>110$), while no clipping was required for the standard and other PD versions of the DSM. 

\begin{figure}
\centerline{\includegraphics[width=0.99\linewidth]{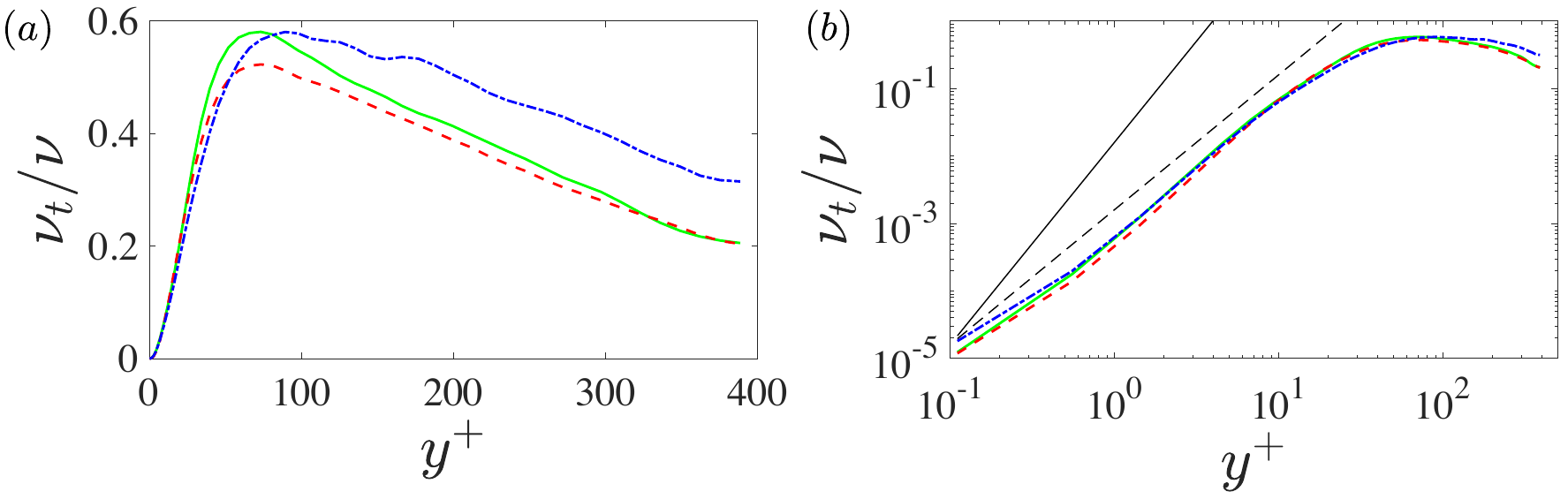}}
\caption{Average SGS eddy viscosity from LES of  channel flow at $Re_{\tau}=395$. SGS eddy viscosity is normalized by kinematic viscosity. ($a$) Linear scale. ($b$) Log scale. Green solid line, LES with DSM; red dashed line, LES with PDL2; blue dash-dotted line, LES with PDWL2. In ($b$),  black solid/dashed lines are $y^3$/$y^2$ reference lines, respectively. 
}
\label{fig:nu}
\end{figure}

\begin{figure}
\centerline{\includegraphics[width=0.5\linewidth]%
{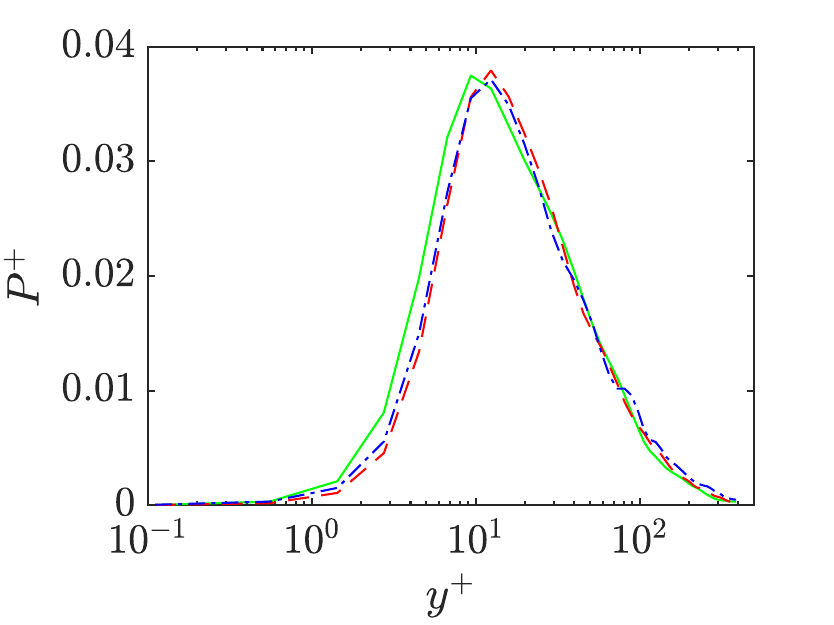}}
\caption{\color{red}Profiles of the SGS energy transfer rate in turbulent channel flow at $Re_{\tau}=395$. $P$ and $y$ are normalized with viscous wall units. Green solid line, LES with DSM; red dashed line, LES with PDL2; blue dash-dotted line, LES with PDWL2.\color{black}}
\label{fig:prod}
\end{figure}

\begin{figure}
\centerline{\includegraphics[width=0.33\linewidth]%
{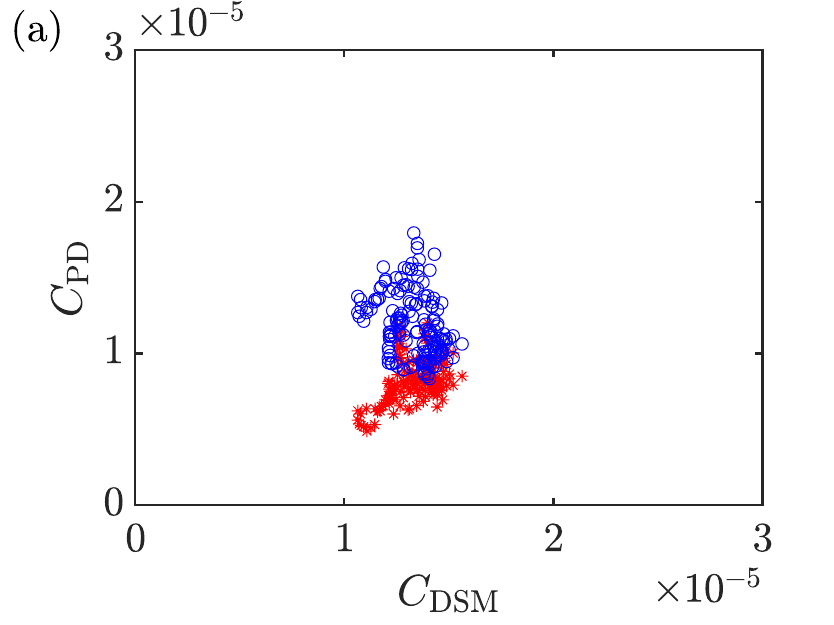}\includegraphics[width=0.33\linewidth]%
{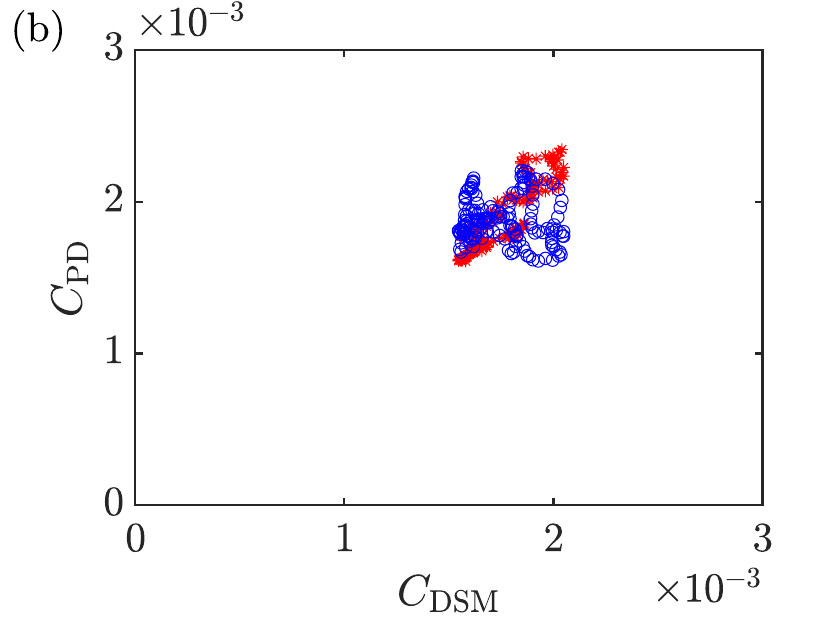}\includegraphics[width=0.33\linewidth]%
{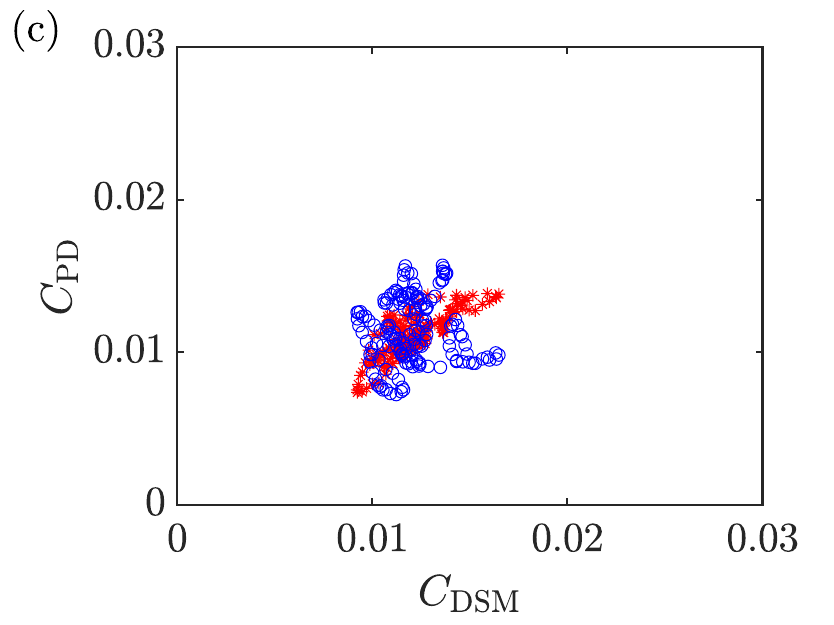}}
\caption{\color{red}Scatter plot of the Smagorinsky coefficients of the dynamic models in turbulent channel flow at $Re_{\tau}=395$. The horizontal axis is the DSM result, and the vertical axis corresponds to the results from the two PD formulations. ($a$) Viscous sublayer ($y^+=1.5$); ($b$) buffer layer ($y^+=15$); ($c$) log layer ($y^+=45$)
Red, LES with PDL2; blue, LES with PDWL2.\color{black}}
\label{fig:Smag_coeff}
\end{figure}

Figure~\ref{fig:nu} shows the time-averaged SGS eddy viscosity across the channel.  The three SGS models produce similar levels of SGS eddy viscosity. The near-wall SGS eddy viscosity exhibits $y^2$ behavior instead of $y^3$,  consistent with the finding of \citet{NomaPark2009} where the SGS eddy viscosity computed from DNS data of channel flow at $Re_{\tau} = 590$ also exhibited $y^2$ behavior near the wall. 
\color{red}
Figure.~\ref{fig:prod} shows the profiles of the SGS energy transfer rate $P=\tau_{ij}\bar{S}_{ij}$ in the turbulent channel flow. This result is directly relevant to the energy transfer between the large and small scales. Again, there is no significant difference among DSM, PDL2 and PDWL2 results, showing that these three formulations are 
largely equivalent in terms of capturing the energy transfer.
The instantaneous Smagorinsky coefficients collected over one eddy turnover time ($t = \delta/u_{\tau}$) are shown in the scatter plots in Fig.~\ref{fig:Smag_coeff}. The initial condition is the converged flow field calculated with DSM. Three different formulations are then applied to calculate the instantaneous Smagorinsky coefficients. Pairings in the scatter plots are such that the data involved are sampled at the same simulation time. 
Three different wall normal locations including viscous sublayer, buffer layer and log layer are probed. In Fig.~\ref{fig:Smag_coeff}, most points are observed to be clustered, lying generally within 1 $\sim $ 2 standard deviations of the data involved.  
The PD formulations produce Smagorinsky coefficients quite close to the those produced by the standard DSM, as expected. The level of collapse is found  higher 
in the buffer layer and log layer than in the  viscous sublayer, 
where the difference in the latter is deemed insignificant because 
the model contribution is negligibly small even compared to the molecular viscosity.
\color{black}
\color{black}

\color{black}

\begin{figure}
\centerline{\includegraphics[width=0.99\linewidth]{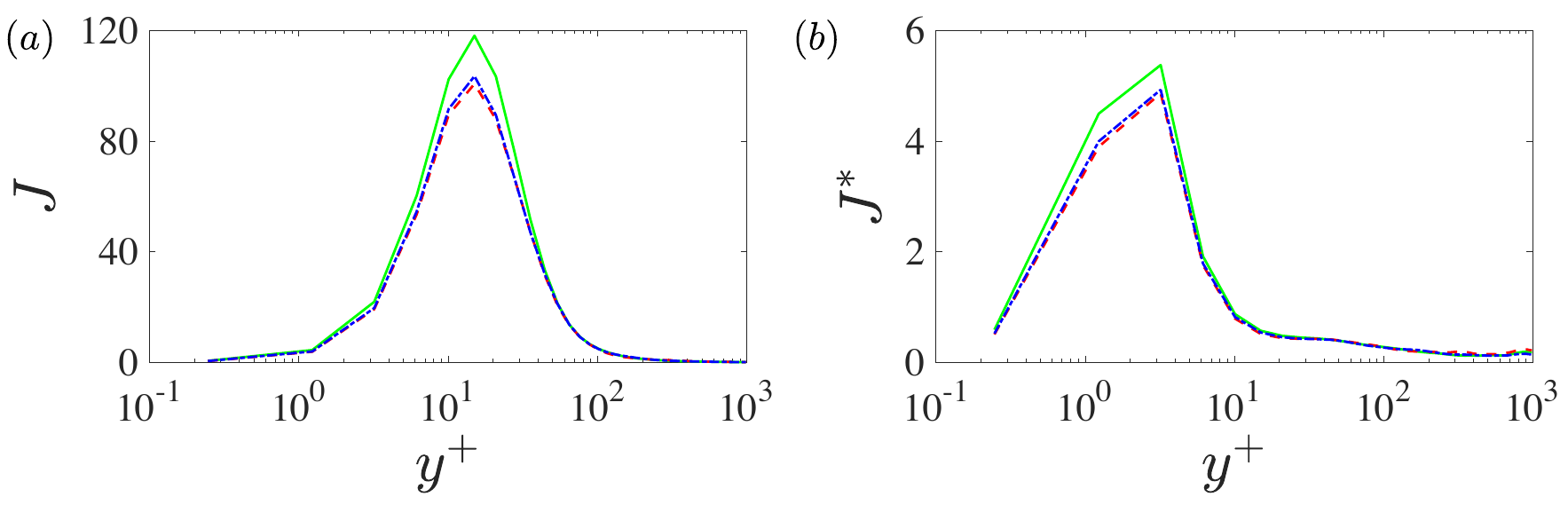}}
\caption{Profile of $J$ (the L2 norm of the GIE tensor) along the wall-normal direction in  channel flow at $Re_{\tau}=1000$. ($a$)  $J$ is not normalized; ($b$)  $J$ is normalized by a combination of mean velocity gradient $dU/dy$ and Reynolds stress $\left<u'u'\right>$. Green solid line, LES with DSM; red dashed line, LES with PDL2; blue dash-dotted line, LES with PDWL2.}
\label{fig:QijQij1000}
\end{figure}

The effects of different model formulations can also be evaluated through the norm of the GIE tensor $Q_{ij}$, given by $J = Q_{ij}Q_{ij}$  
($Q_{ij}$  defined in Eq.~(\ref{eq:Germano error})). 
 \citet{NomaPark2009} and \citet{Toosi2021} pointed out that the GIE will be zero for the exact SGS model, and that a good SGS model should pursue small GIE.
We focus on the coarse LES case of $Re_{\tau}=1000$, but the same trend is observed in the  $Re_{\tau}=395$ case as well.  
The profile of $J$ in Fig.~\ref{fig:QijQij1000}($a$) shows that 
the peak location of the GIE is at around $y^+=10$ within the buffer layer, consistent with findings of \citet{NomaPark2009}. 
It is found that the GIE from the original DSM is almost identical to the GIE from the PD formulations, except in the buffer layer ($y^+ = 5 \sim 30$). In the buffer layer, about 15\% reduction in the peak GIE is observed with the PD formulations as compared to the original DSM. 
Figure~\ref{fig:QijQij1000}($b$) presents the normalized $J$ profile. Here,  $J$ is normalized by $\left(\frac{d\left<U\right>}{dy}\delta_{\nu}\right)^4+\left<u'u'\right>^2$, 
where $\delta_\nu = \nu / u_{\tau}$. This normalization includes the strain rate $\frac{d\left<U\right>}{dy}$ and Reynolds stress $\left<u'u'\right>$ which are related to $M_{ij}$ and $L_{ij}$ in the GIE tensor. 
This normalization produces $J = O(1)$, suggesting 
that the mixed viscous/turbulent scaling is effective for the GIE. 
Under such normalization, the peak error appears around 
$y^+ =5$. 
\color{red}
The PD formulations are derived based on the Germano identity. In general, PD formulation can be constructed for any dynamic model based on a similar Germano identity. In Appendix \ref{app_Vreman}, the PDL2 formulation is applied to the dynamic Vreman model of \citet{Lee2010dynamic}, where we find its result is similar to what has been presented for DSM in this section. 
\color{black}

\subsection{Three-dimensional turbulent boundary layer}
\begin{figure}
\centerline{\includegraphics[width=1.0\linewidth]{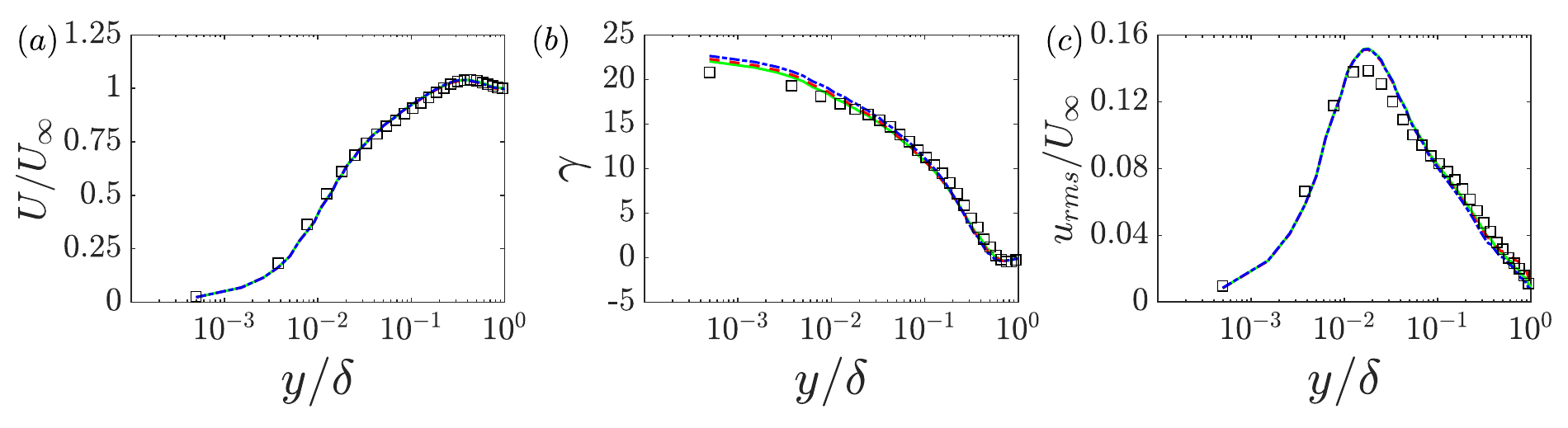}}
\caption{Mean velocity in the 3DTBL. ($a$) Mean velocity magnitude; ($b$) Mean flow direction $\gamma = \arctan(W/U)$; ($c$) \color{black} Free-stream-wise turbulence intensity, $u_{rms}$\color{black}. Squares, DNS \citep{Spalart1989}; green solid line, LES with DSM; red dashed line, LES with PDL2; blue dash-dotted line, LES with PDWL2.}
\label{fig:3dtbl_pdl2}
\end{figure}

The idea of the reduced dynamic procedure in the DSM is also examined in a 3DTBL. In this flow, the freestream velocity vector is rotating at a constant angular velocity. The flow is 
statistically steady 
in the coordinate system rotating with the freestream. $x$ and $z$ 
denote the directions parallel/perpendicular to the freestream, respectively. 
Figure~\ref{fig:3dtbl_pdl2}($a$) shows the mean velocity magnitude profile in the 3DTBL. For the velocity magnitude, the three different formulations produce almost identical results, showing reasonable agreement with the DNS. 
A salient feature of 3DTBLs is the variation of the flow direction with wall distance. 
The mean flow direction is quantified  
in Fig.~\ref{fig:3dtbl_pdl2}($b$) 
using the 
flow angle in wall-parallel planes, 
$\gamma = \arctan(W/U)$, 
where $U$ and $W$ are aligned with/perpendicular to the freestream, respectively. 
The agreement with DNS is slightly worse compared to that of the velocity magnitude. LES solutions have about 3 degrees of discrepancy close to the wall, and a slight underprediction of the flow angle is seen in the outer layer. 
PDL2 and DSM produce nearly identical predictions. PDWL2 is relatively worse in $y/\delta < 0.1$, but the agreement is still reasonable. 
Overall, the two modified PD models are as good as the original DSM. \color{black}
The free-stream-wise turbulence intensity is shown in Fig.~\ref{fig:3dtbl_pdl2}($c$). Similar to the mean flow statistics, there is negligible difference among three formulations of the dynamic models.\color{black}

In the Appendix \ref{appB}, the reduced dynamic procedure is also applied to a non-Boussinesq tensor-coefficient SGS model in the same flow. This type of model is better suited for 3DTBLs, because the stress/strain alignment assumption in the Boussinesq eddy viscosity models is invalid in 3DTBLs. 
Overall, the PD formulation shows slightly improved performance compared to the original dynamic tensor-coefficient SGS model of \citet{Agrawal2022}. Details related to this aspect can be found in the Appendix \ref{appB}. 

\subsection{Flow over periodic hills}
\begin{figure}
\centerline{\includegraphics[width=1\linewidth]{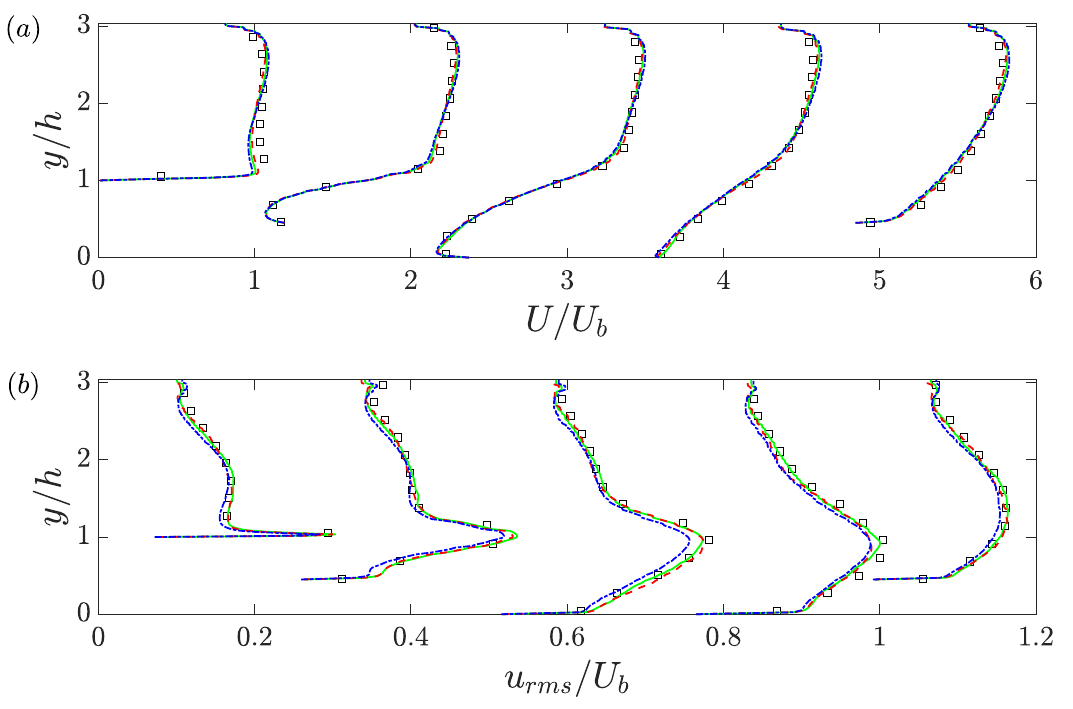}}
\caption{($a$)Streamwise mean velocity profiles in the flow over periodic hills ($Re = 10595$) from $x/h = 0.05,\ 1.00,\ 2.00,\ 4.00,\ 8.00$. Profiles are shifted along the abscissa by 1.2; ($b$)  \color{black}Streamwise turbulence intensity profiles at the same 5 stations. Profiles are shifted along the abscissa by 0.24\color{black}. Black squares, experiment \citep{Rapp2011}; green solid line, LES with DSM; red dashed line, LES with PDL2; blue dash-dotted line, LES with PDWL2.}
\label{fig:periodic_hill}
\end{figure}

\begin{figure}
\centerline{\includegraphics[width=0.5\linewidth]%
{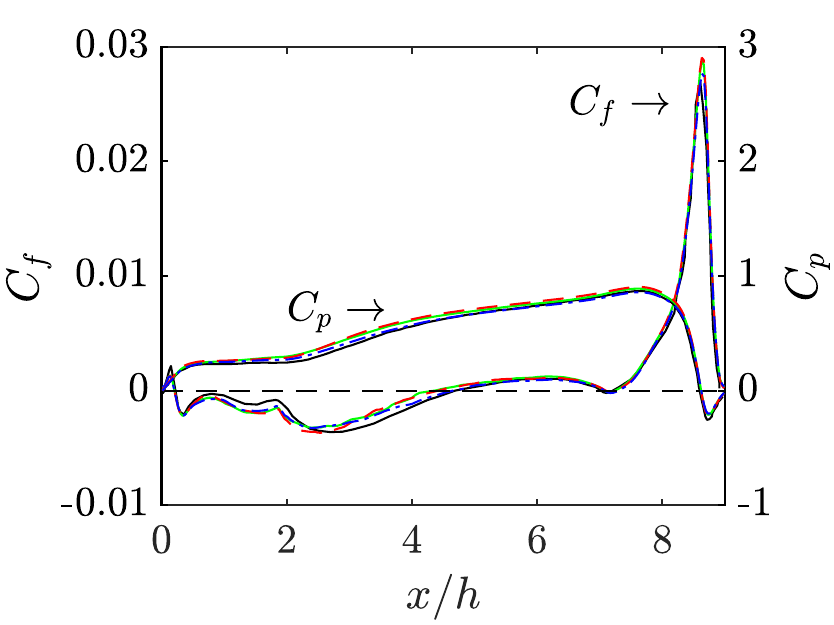}}
\caption{\color{red}Skin friction and pressure coefficient distribution along the bottom wall of the periodic hill.  Green solid line, LES with DSM; red dashed line, LES with PDL2; blue dash-dotted line, LES with PDWL2; black dashed line, LES from \cite{Frohlich2005highly}.\color{black}}
\label{fig:cf_cp_hill}
\end{figure}

\begin{figure}
\centerline{\includegraphics[width=1.0\linewidth]%
{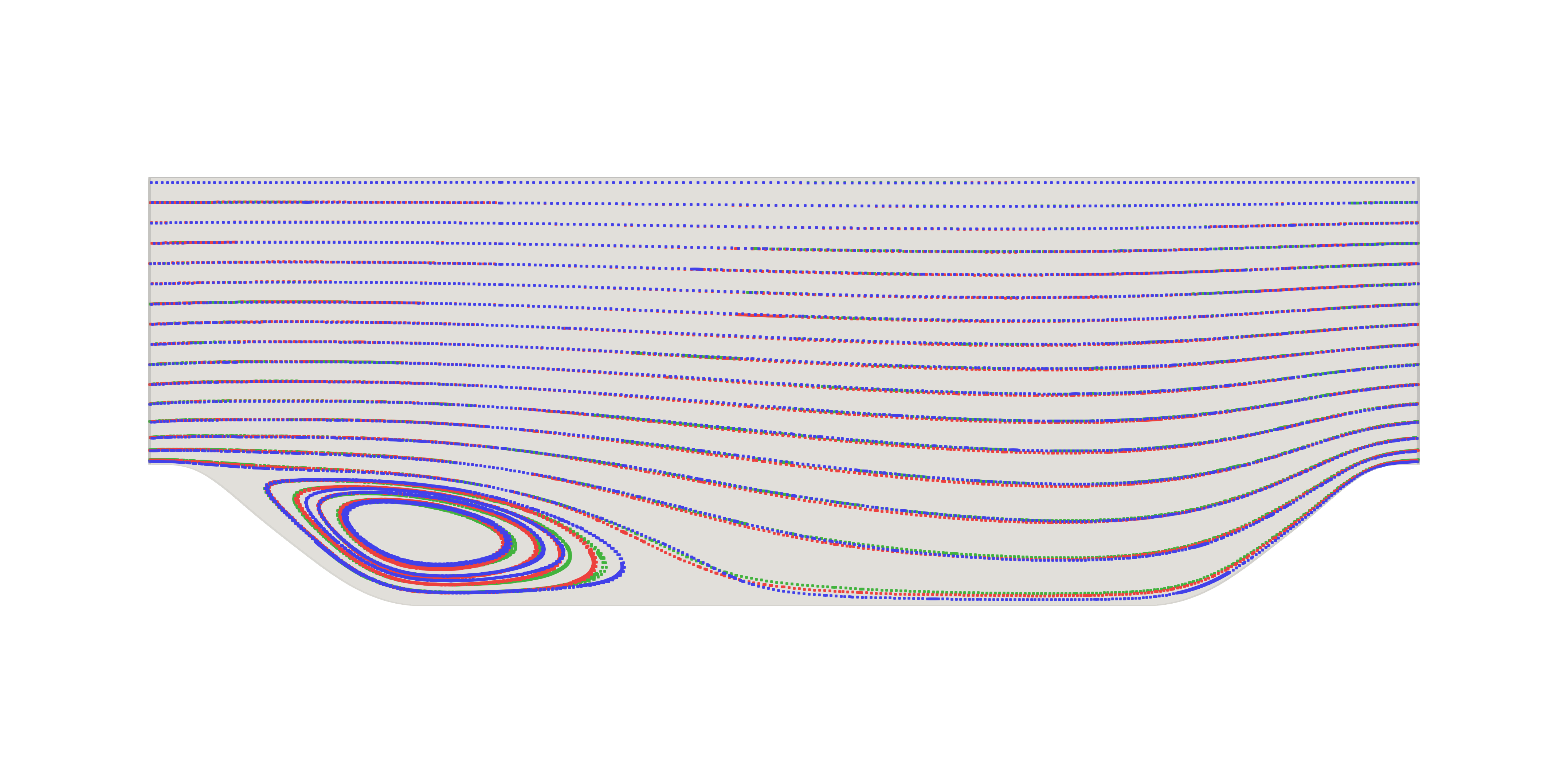}}
\caption{\color{red}Streamlines of the flow over periodic hills.  Green, LES with DSM; red, LES with PDL2; blue, LES with PDWL2.\color{black}}
\label{fig:streamline}
\end{figure}

Figure~\ref{fig:periodic_hill}($a$) shows the 
the mean streamwise velocity in the separating flow over periodic hills predicted with 
the DSM with different dynamic procedures. 
 Good agreements is found with the experiment in all three SGS models.  The largest discrepancy is observed at $x/h = 0.05$ close to the separation point  ($x/h \approx 0.2$). 
 At this location, PDL2 shows slightly better performance than the other two formulations.
 Overall, it is found that the reduced PD formulations perform as good as the original DSM. \color{black}Figure~\ref{fig:periodic_hill}($b$) presents the turbulence intensity profiles at the same 5 stations. The LES results agree reasonably well with the experiment. The DSM and PDL2 show slightly better prediction of the peak value at $x/h=2.00$ compared to PDWL2. \color{black}
 \color{red}
 The skin friction coefficient and pressure coefficient distributions are shown in Fig.~\ref{fig:cf_cp_hill}. $C_f$ and $C_p$ results are almost identical among three formulations and agree well with the reference LES results \cite{Frohlich2005highly}. In Fig.~\ref{fig:streamline}, the streamlines of the periodic hill case are presented. The three formulations produce almost the same results. In the separated flow region, the separation bubble size is slightly larger in PDWL2 result.
 \color{black}
 The overall performance are almost equivalent among three formulations.
 \color{black}

\color{black}

\color{black}

\section{\label{sec:conclusion}Conclusion}
Motivated from vorticity dynamics, a hidden mechanism at work in the success of dynamic LES SGS models is explored. 
Based on the assumption that the SGS eddies tend to be aligned with the principal stretching/compression directions of the resolved flow field, we postulate that only a few elements 
of the Germano identity, pertaining to the principal directions of the resolved strain-rate, matter in the dynamic procedure to determine the model coefficient. 
Some principal-direction (PD) variants of the DSM based on this idea are tested in canonical turbulent channel flows, a three-dimensional turbulent boundary layer, and a separating flow over periodic hills. In all the cases, PD formulations produced almost identical results as the original DSM. 
These results demonstrate that not all components of the Germano identity matters, and that 
satisfaction of the Germano identity 
along some of the principal directions of the resolved strain-rate tensor might be the essence of the dynamic procedure.  
This establishes a physical connection between the Germano identity, initially perceived as a purely mathematical identity devoid of physics, and vorticity in the resolved flow field. This connection provides an insight into why dynamic models succeed and offers guidance for future efforts in subgrid-scale modeling.


\begin{acknowledgments}
This research was sponsored by NASA's Transformational Tools and Technologies Project of the Transformative Aeronautics Concepts Program under the Aeronautics Research Mission Directorate (Grant 80NSSC18M0155). Computational resources supporting this work were provided by the NASA High-End Computing Program through the NASA Advanced Supercomputing Division at Ames Research Center.
\end{acknowledgments}

\appendix
\section{Application to dynamic Vreman model}\label{app_Vreman}
\color{red}
Reduction of the dynamic procedure onto the principal directions of the strain rate tensor can be applied to the dynamic Vreman model (DVM) of \cite{Lee2010dynamic}. 
In the original DVM, the SGS stresses are modeled as 

\begin{equation} \label{eq:DVM}
    \tau_{ij}-\frac{\tau_{kk}}{3}\delta_{ij} =2C_{\nu}\sqrt{\frac{\Pi_{\overline{\beta}}}{\overline{\alpha}_{ij}\overline{\alpha}_{ij}}}\overline{S}_{ij}, 
\end{equation}
where
\begin{equation} \label{eq:DVM_alpha}
    \alpha_{ij} = \frac{\partial u_j}{\partial x_i},
\end{equation}
\begin{equation} \label{eq:DVM_pi}
    \Pi_{\overline{\beta}} = \overline{\beta}_{11}\overline{\beta}_{22}+\overline{\beta}_{22}\overline{\beta}_{33}+\overline{\beta}_{33}\overline{\beta}_{11}-\overline{\beta}^2_{12}-\overline{\beta}^2_{23}-\overline{\beta}^2_{31},
\end{equation}
\begin{equation} \label{eq:DVM_beta}
    \overline{\beta}_{ij} = \sum_{m=1}^3\Delta^2_{m}\overline{\alpha}_{mi}\overline{\alpha}_{mj}.
\end{equation}
Here, $C_{\nu}$ is the Vreman model coefficient and $\Delta_m$ is the characteristic filter width in the m$^{th}$ direction.
The unknown coefficient in the original DVM is determined by 
minimizing the GIE over the whole computational domain, resulting in the model coefficient which is a function of time only. 
We proceed with a manner similar to DSM, using  the Germano identity: 
\begin{equation} \label{eq:Germano_DVM}
    L_{ij} - \frac{1}{3}\delta_{ij}L_{kk} = C_{\nu}M_{ij}, 
\end{equation}
where
\begin{equation} \label{eq:mij_DVM}
    M_{ij} = 2\sqrt{\frac{\Pi_{\widehat{\overline{\beta}}}}{\widehat{\overline{\alpha}}_{ij}\widehat{\overline{\alpha}}_{ij}}}\widehat{\overline{S}}_{ij}-2\widehat{\sqrt{\frac{\Pi_{\overline{\beta}}}{\overline{\alpha}_{ij}\overline{\alpha}_{ij}}}\overline{S}_{ij}}
\end{equation}
and $L_{ij}$ is the same as in Eq.~\ref{eq:lij}.
The unknown coefficient can be calculated as 
\begin{equation} \label{eq:dvm_C}
    C_{\nu} = \left < L_{ij}M_{ij}\right >_{V} 
    \Big / 
    \left < M_{ij}M_{ij}\right >_{V},   
\end{equation}
where $\left<\cdot\right>_{V}$ denotes the instantaneous volume averaging over the entire computational domain.

\begin{figure}
\centerline{\includegraphics[width=1.0\linewidth]%
{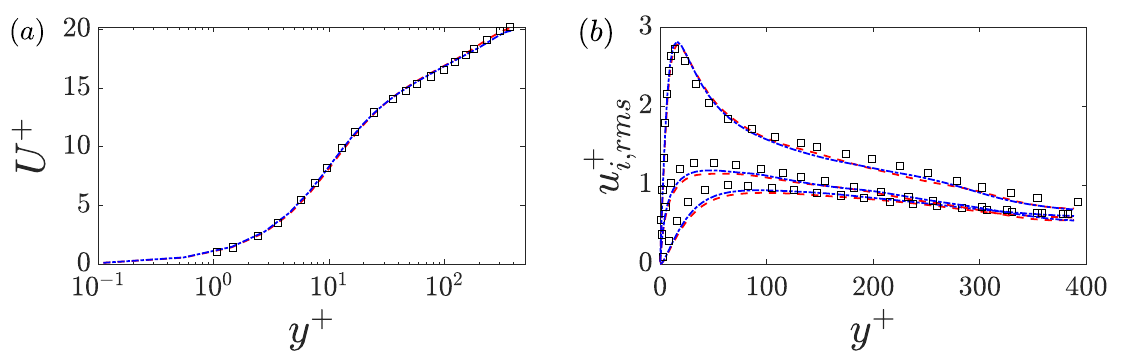}}
\caption{\color{red}Profiles of flow statistics in wall units in turbulent channel flow at $Re_{\tau}=395$. ($a$) Mean streamwise velocity; ($b$) Turbulence intensities ($u_{rms}$, $v_{rms}$ and $w_{rms}$). Black squares, DNS \cite{Moser1999}; red dashed line, LES with DVM; blue dash-dotted line, LES with DVM-PD.\color{black}}
\label{fig:pd_vreman}
\end{figure}

The PDL2 formulation of DVM (referred to as DVM-PD)
can be constructed in the same manner as discussed earlier for DSM. 
This involves modifying Eq. (A7) into the form of 
Eq.~(\ref{eq:pd_general}), utilizing the information pertaining only to the eigen directions of the resolve strain rate. 
This extension to DVM is straightforward, 
 because both DSM and DVM are based on the Germano identiy, serving as the foundation for any PD formulations derived from them. 
DVM-PD is tested in the turbulent channel flow and the results are shown in Fig.~\ref{fig:pd_vreman}. The DVM and the DVM-PD are almost identical in terms of the mean velocity. Slight difference can be observed for Reynolds stresses but it is almost negligible. 
\\ 
\color{black}

\section{Application to dynamic tensor coefficient Smagorinsky model}\label{appB}
Reduction of the dynamic procedure onto  the principal directions of the strain rate tensor 
can be applied to the dynamic tensor coefficient Smagorinsky model (DTCSM) \citep{Agrawal2022} as well. 
In the original work, the DTCSM models the SGS stress as 
\begin{equation} \label{eq:DTCSM}
    \tau_{ij}-\frac{\tau_{kk}}{3}\delta_{ij} = (C_{ik}S_{kj}+C_{jk}S_{ki})|S|\Delta^2, 
\end{equation}
where $C_{ij}$ is the tensor of model coefficients. 
For the DTCSM, the Germano identity produces 
\begin{equation} \label{eq:DTCSM-GIE}
    L_{ij} = \left( C_{ik}\Delta^2M_{kj}+C_{jk}\Delta^2M_{ki}\right).   
\end{equation}
\citet{Agrawal2022} imposed the trace-free requirement on the model leading to 
the following constraints, 
\begin{equation} \label{eq:DTCSM-coeff}
    C_{11}=C_{22}=C_{33}; \  \ C_{ij} = -C_{ji} (i\neq j), 
\end{equation}
and the 4 independent coefficients were determined 
to best satisfy 6 constraints from the GI in a L$_{2}$  sense. 
Similar to the formulation described in Sec.~\ref{sec:mod},
the GIE tensor $Q_{ij} = L_{ij} - \left( C_{ik}\Delta^2M_{kj}+C_{jk}\Delta^2M_{ki}\right)$ for the DTCSM can be transformed into the principal coordinate system of the filtered strain-rate tensor ($\overline{S}_{ij}$). 
The PD-version of the DTCSM then determines model coefficients by enforcing the GI along the principal directions of  $\overline{S}_{ij}$ only (the diagonal components of the transformed GIE tensor). 
\begin{equation} \label{eq:dtcsm_matrix}
\begin{pmatrix}
L'_{11}\\L'_{22}
\end{pmatrix}
=
\begin{pmatrix}
2M'_{11} & 2M'_{12} & 2M'_{13} & 0\\
2M'_{22} & -2M'_{12} & 0 & 2M'_{23}
\end{pmatrix}
\begin{pmatrix}
C_{11}\\C_{12}\\C_{13}\\C_{23}
\end{pmatrix}.
\end{equation}
It can observed readily that there are 4 unknown variables and but only two constraints. 
To close the system, 
we introduce an additional assumption on the coefficients operating on the non-principal components of $M_{ij}$, 
letting $C_{12} = C_{13} = C_{23}$. This leads to a closed 2x2 system from which the model coefficients can be determined. 
\begin{equation} \label{eq:dtcsm_matrix_reduced}
\begin{pmatrix}
L'_{11}\\L'_{22}
\end{pmatrix}
=
\begin{pmatrix}
2M'_{11} & 2M'_{12}+2M'_{13}\\
2M'_{22} & -2M'_{12}+2M'_{23}
\end{pmatrix}
\begin{pmatrix}
C_{11}\\C_{12}
\end{pmatrix}.
\end{equation}

\begin{figure}
\centerline{\includegraphics[width=1\linewidth]{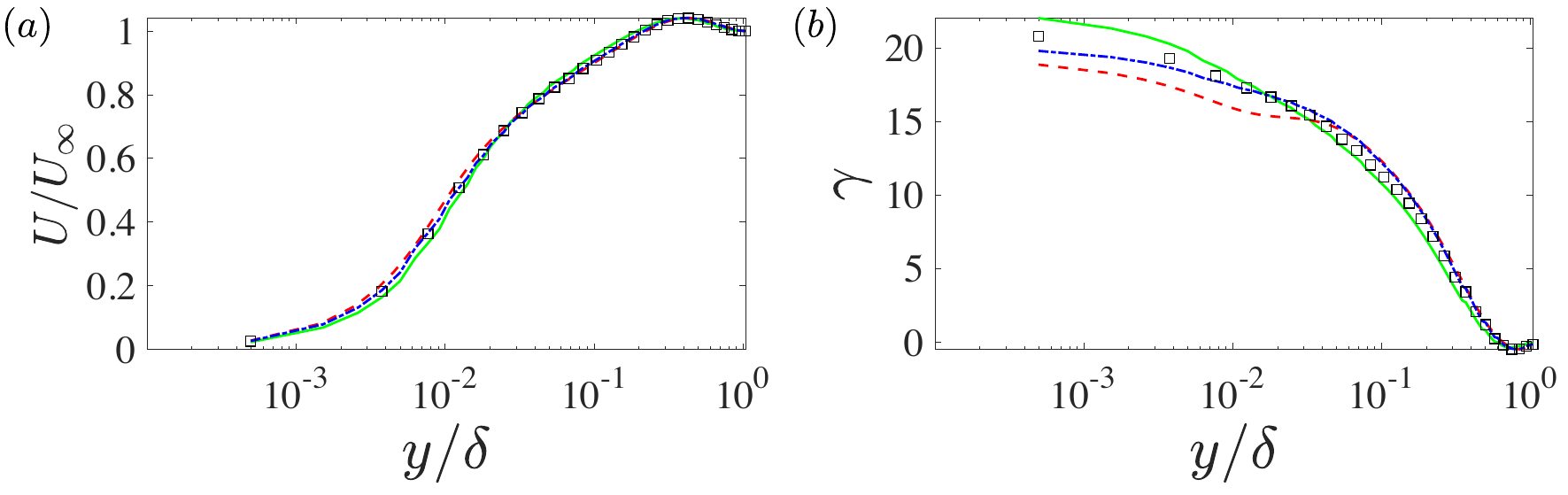}}
\caption{Mean velocity in the 3DTBL. ($a$) Mean velocity magnitude; ($b$) Mean flow direction $\gamma = \arctan(W/U)$. 
Black squares, DNS; green solid line, LES with DSM; red dashed line, LES with DTCSM; blue dash-dotted line, LES with DTCSM-PD.}
\label{fig:3dtbl_dtcsm}
\end{figure}

The DTCSM and its PD-variant (denoted here as DTCSM-PD) is applied to the 3DTBL case considered in Sec.~\ref{sec:result}. 
Figure~\ref{fig:3dtbl_dtcsm} shows the profiles of the mean velocity magnitude and flow direction. For the velocity magnitude, the two tensor-coefficient SGS models have a very good agreement with the DNS, while the DSM has a slight discrepancy (under/over prediction in the near-wall/bulk regions). For the flow direction, it is clear that the DTCSM-PD has the best performance close to the wall, where the DSM overpredicts and the DTCSM undepredicts the flow angle. For $y/\delta>0.2$, the DTCSM and the DTCSM-PD produce almost identical results for the flow direction and they  agree well with the DNS. The original DSM slightly underpredicts the flow angle at $y/\delta>0.2$.
Overall, we again confirm that reduction of the dynamic procedure along the principal direction is as effective as the original DTCSM. In fact, 
 the DTCSM-PD has the best prediction of the mean velocity. 
We do note that the choice of the additional constraints ($C_{12} = C_{13} = C_{23}$) is somewhat arbitrary. Other choices are possible, and how we close the system may 
 affect the performance of the model. The purpose of this appendix is to provide a potential extension of the conclusion in the main text towards more comprehensive SGS models.

\nocite{*}

\bibliography{apssamp}

\end{document}